\documentclass{jfm}
\usepackage{graphicx}
\usepackage{epstopdf, epsfig}

\usepackage{natbib}
\usepackage{lmodern}
\usepackage[utf8]{inputenc}
\usepackage[english]{babel}
\usepackage[T1]{fontenc}
\usepackage{amsmath,bm}
\usepackage{amsfonts}
\usepackage{amssymb}
\usepackage[hidelinks]{hyperref}
\usepackage{eurosym}
\usepackage{pst-all} 
\usepackage{subcaption} 
\usepackage{empheq}
\usepackage{enumerate}
\usepackage{comment}
\usepackage{multirow}
\usepackage[normalem]{ulem} 

\def\la{\left \langle}
\def\ra{\right \rangle}

\title{Rapidly rotating radiatively driven convection: experimental and numerical validation of the `geostrophic turbulence' scaling predictions}

\author{Gabriel Hadjerci\aff{1}
  \corresp{\email{gabriel.hadjerci@cea.fr}},
  Vincent Bouillaut\aff{2}, Benjamin Miquel\aff{3}
 \and Basile Gallet\aff{1}}

\affiliation{\aff{1}Université Paris-Saclay, CNRS, CEA, 91191, Gif-sur-Yvette, France
\aff{2}DAAA, ONERA, Université Paris-Saclay F-92322 Châtillon, France
\aff{3}CNRS, Ecole Centrale de Lyon, INSA Lyon, Universite Claude Bernard Lyon 1, LMFA, UMR5509, 69130 Ecully}

\begin{document}

\maketitle

\begin{abstract}

We experimentally and numerically characterize rapidly rotating radiatively driven thermal convection, beyond the sole heat transport measurements reported by \cite{bouillaut_experimental_2021}. Based on a suite of direct numerical simulations (DNS) and additional processing of the experimental data collected by \cite{bouillaut_experimental_2021}, we report the simultaneous validation of the scaling predictions of the `geostrophic turbulence' regime -- the diffusivity-free or `ultimate' regime of rapidly rotating convection -- for the heat transport and the temperature fluctuations. {Following such cross-validation between DNS and laboratory experiments, we further process the numerical data to validate the `geostrophic turbulence' scaling predictions for the flow velocity and horizontal scale.} Radiatively driven convection thus appears as a versatile setup for the laboratory observation of the diffusivity-free regimes of various convective flows of geophysical and/or astrophysical interest.

\end{abstract}

\section{Introduction}

Thermal convection is a key process driving natural turbulent flows, be it in stellar and planetary interiors or in the open ocean \citep{stevenson_turbulent_1979,marshall_openocean_1999,aurnou2015rotating,de2016strong,hindman2020morphological,vasil2021rotation}. In these various contexts, global rotation strongly affects the resulting flow and its transport properties through the action of the Coriolis force \citep{ecke2023turbulent}. The {Reynolds number takes prohibitively} large values in natural settings, beyond direct reach of laboratory and numerical models. The standard approach thus consists in inferring scaling-laws based on dimensional analysis, asymptotic theory, experimental studies and numerical studies, with the goal of extending these scaling-laws to the extreme parameter values that characterize natural flows.

The scaling relation between the convective heat flux and the overall temperature drop has received most attention, as it characterizes both the heat transport properties of the system and the energetics of the flow. When it comes to the modeling of natural flows, a general belief is that the tiny molecular values of the thermal and momentum diffusivities of the system should not enter this scaling relation \citep{spiegel_convection_1971,marshall_openocean_1999}. Such a `diffusivity-free' argument can be traced back to the zeroth-law of turbulence, understood in a broad sense: in a fully turbulent system, the large-scale quantities {should be} related in a way that does not involve the tiny molecular diffusivities.

For non-rotating convection, the diffusivity-free argument alone allows {one} to conclude on the scaling relation between the dimensionless heat flux (the Nusselt number $Nu$) and the dimensionless temperature drop (the temperature-based Rayleigh number $Ra^{(\Delta T)}$). This leads to the so-called `ultimate' scaling regime of thermal convection \citep{kraichnan_turbulent_1962,spiegel_generalization_1963}. By contrast, rotating convection involves an additional dimensionless parameter (the Ekman number $E$) and the diffusivity-free argument alone does not suffice to predict the scaling behavior of the system. \cite{stevenson_turbulent_1979} therefore introduced additional scaling arguments, {many of which were inferred (perhaps surprisingly) from} the linear stability analysis of rapidly rotating convection. {Combining these additional arguments with the diffusivity-free assumption, \cite{stevenson_turbulent_1979} obtained definite}  predictions for the scaling behavior of the main quantities of interest: heat transport, temperature fluctuations, flow speed and characteristic scales. 
{A more mathematically-rooted way to arrive at the same scaling predictions stems from the work of~\cite{julien_new_1998}, who carried out an asymptotic expansion of the governing equations in the rapidly rotating regime. }
This procedure results in a reduced set of equations that involves fewer dimensionless control parameters: instead of the Rayleigh number $Ra^{(\Delta T)}$ and the Ekman number $E$ arising independently in the equations, the reduced set of equations involves only a reduced Rayleigh number $\widetilde{Ra}=Ra^{(\Delta T)} E^{4/3}$. 
{Provided one starts from the reduced set of equations of~\cite{julien_new_1998}, the sole diffusivity-free assumption yields all the scaling predictions of \cite{stevenson_turbulent_1979}, confirming Stevenson's impressive physical insight. }
The corresponding fully turbulent {scaling} regime of rapidly rotating convection is referred to as the `Geostrophic Turbulence' (GT) scaling regime in the more recent literature \citep{chengGAFD18}.

An ongoing line of research consists in trying to reach the GT regime using Direct Numerical Simulation (DNS). This approach has proven successful in a range of idealized models of increasing numerical complexity. DNS of the reduced equations \citep{julien_heat_2012,plumley_effects_2016} point to the GT regime at large reduced Rayleigh number. DNS of the full Boussinesq equations {for the Rayleigh-Bénard setup with free-slip boundary conditions} also point to the GT scaling regime for the heat transport \citep{stellmach_approaching_2014}. Departing from the standard Rayleigh-Bénard setup, \cite{barker_theory_2014} introduced internal heating and cooling in the vicinity of the lower and upper boundaries, respectively, and validated the various scaling predictions of the GT regime (see also \cite{currie2020convection}). More recently, \cite{song2024scaling} have reported DNS of Rayleigh-B\'enard convection with no-slip boundaries ; the heat transport scaling-law appears to be compatible with the GT prediction for the lowest value of the Ekman number considered in their study, while the GT scaling predictions for the flow speed and flow structure appear to be satisfied over a more extended region of parameter space.

At the experimental level, there are challenging constraints to observe the GT regime in the laboratory. Achieving high Rayleigh number and low Ekman number suggests building tall Rayleigh-B\'enard cells while keeping the radius of the tank small enough to avoid centrifugal effects \citep{ecke_heat_2014,cheng_laboratory-numerical_2015,chengGAFD18,cheng_laboratory_2020,zhang_boundary_2020,wedi_rotating_2021}. An issue with such cigar-shaped containers is the emergence of wall-modes and boundary zonal flows that can contribute significantly to the overall heat transport \citep{favier2020robust,zhang_boundary_2021,wedi_experimental_2022}. It is therefore unclear whether the GT regime can be achieved in the current generation of rotating Rayleigh-B\'enard experiments, and determining the optimal laboratory design that could potentially lead to the observation of the GT regime in rotating Rayleigh-B\'enard convection remains an intense topic of ongoing research \citep{chengGAFD18,kunnen2021geostrophic,terrien_suppression_2023}.

Departing from the standard Rayleigh-B\'enard setup, we have introduced an experimental apparatus where convection is driven radiatively, through the absorption of visible light by a dyed fluid. Through a combination of radiative heating and effective internal cooling, one can bypass the top and bottom boundary layers and observe regimes of thermal convection that are controlled by the bulk turbulent flow, such as the `ultimate' regime of (non-rotating) turbulent convection \citep{lepot_radiative_2018,bouillaut_transition_2019,miquel_convection_2019,miquel_role_2020}. Global rotation was recently added {to} the laboratory setup, providing the first experimental observation of the GT heat transport scaling relation \citep{bouillaut_experimental_2021}. {This experimental configuration shares many similarities with the numerical setup introduced by \cite{barker_theory_2014} (see also \cite{currie2020convection}), differing primarily in the distribution of internal heat sources and sinks, in the boundary conditions, and in the precise definitions of the diagnostic variables.} The goal of the present study is to characterize the turbulent state of radiatively driven rotating convection beyond the sole heat transport, and assess the validity of the GT scaling predictions for the temperature fluctuations, the flow speed and the characteristic scale of the flow. One issue at the experimental level is that the fluid is opaque, making global velocity measurements challenging (see \cite{bouillaut_velocity-informed_2022} for velocity estimates in the non-rotating experiment). We thus carried out a hybrid experimental-numerical study of radiatively driven convection combining further processing of the temperature measurements in \cite{bouillaut_experimental_2021} with in-depth numerical diagnostics of the speed and structure of the flow. We validate the numerical approach as a good model of the system by intercomparison of the experimental and numerical data for temperature, before leveraging the 3D DNS to characterize the velocity field.

\section{Radiatively driven rotating convection}

\begin{figure}
    \centering
    \includegraphics[width=14cm]{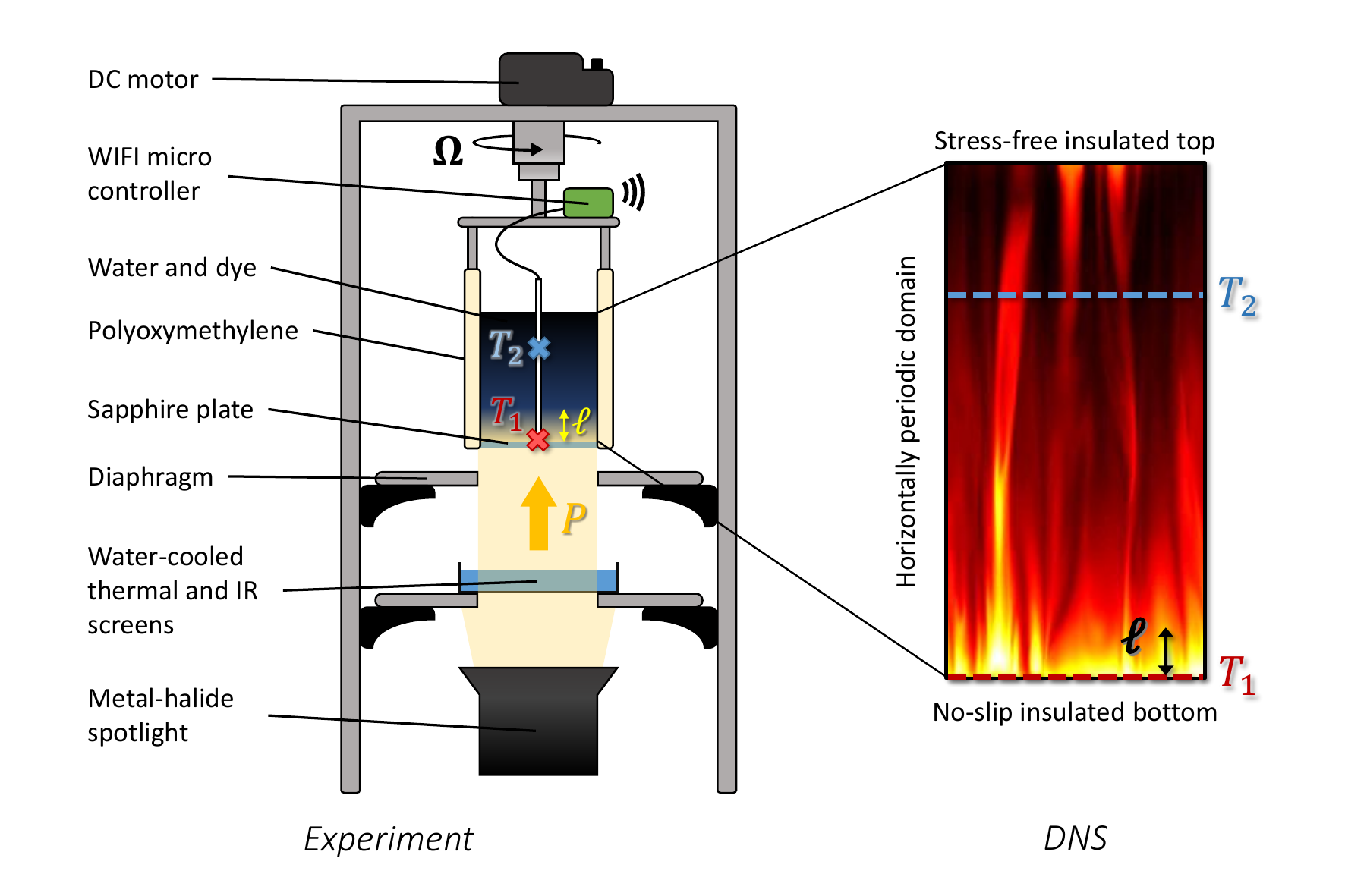}
    \caption{Experimental and (left) numerical (right) implementations of radiatively driven rotating convection. Absorption of an upward flux of light by the dyed fluid induces an internal heat source that decreases exponentially with height measured from the bottom of the fluid domain, over an absorption length $\ell$. Secular heating of the fluid induces uniform effective internal cooling compensating the radiative heat source on vertical average.}
    \label{fig:experiment_and_DNS}
\end{figure}

\subsection{Theoretical setup}\label{sec:equations}

We consider a horizontal fluid layer of height $H$ radiatively heated from below and thermally insulated at all boundaries (see Figure~\ref{fig:experiment_and_DNS} for a schematic of the laboratory implementation). The fluid absorbs light at the same rate for all incoming wavelengths, which leads to a typical absorption length $\ell$. Following Beer-Lambert's law and denoting as $P$ the radiative flux (per unit surface) impinging on the bottom boundary $z=0$ of the fluid layer, the absorption of light induces an internal heat source that decreases exponentially with height $z$:
\begin{equation}
    Q_H(z)=\frac{P}{\ell} e^{-z/\ell} \, . \label{eq:defQH}
\end{equation}
Together with such radiative heating, we consider a uniform internal cooling term $Q_C=-(1-e^{-H/\ell})P/H < 0$. The volume integral of this cooling term over the fluid domain is opposite to that of the heating term. That is, the overall heat input by the radiative heat source is exactly balanced by the overall heat removed by the uniform heat sink, which ensures a statistically stationary state for the temperature and velocity fields (see section~\ref{sec:labimp} for the experimental implementation of the effective cooling term). 

The fluid layer rotates at a rate $\Omega \mathbf{e}_z$ with respect to an inertial frame and is subject to gravity $-g\mathbf{e}_z$. We restrict attention to the range of parameters where the centrifugal acceleration is negligible \citep{horn_regimes_2018,horn_rotating_2019}. Following the Boussinesq approximation \citep{spiegel_boussinesq_1960}, the fluid properties are assumed to be constant and uniform, with the exception of the density $\rho$, whose variations are retained in the buoyancy force only and are assumed to vary linearly with temperature. The velocity field is thus divergence-free. Denoting the reduced pressure field as $p({\bf x},t)$, the temperature variable as $\theta({\bf x},t)$ and the velocity field in the rotating frame as $\mathbf{u}({\bf x},t)$, the governing equations read:
\begin{eqnarray}
    \partial_t\mathbf{u} + (\mathbf{u}\cdot \bm{\nabla})\mathbf{u} +2\Omega \mathbf{e}_z \times \mathbf{u} &=& -\bm{\nabla}p + \alpha g \theta \mathbf{e}_z + \nu \bm{\nabla}^2\mathbf{u} \, , \label{eq:momentum_eq}  \\
    \nabla\cdot\mathbf{u}&=&0 \, , \label{eq:div_free} \\
    \partial_t \theta + \mathbf{u}\cdot \bm{\nabla}\theta &=& \kappa \bm{\nabla}^2 \theta + \frac{Q_H(z)+Q_C}{\rho C} \, , \label{eq:temperature_eq}  
\end{eqnarray}
where $C$ denotes the specific heat capacity of the fluid, $\alpha$ the coefficient of thermal expansion, $\rho$ the reference density, $\kappa$ the thermal diffusivity and $\nu$ the kinematic viscosity.

The set of equations above involves four dimensionless control parameters:
\begin{align}
    Ra^{(P)}=\frac{\alpha g P H^4}{\rho C \kappa^2 \nu},&& E=\frac{\nu}{{2}\Omega H^2},&& Pr=\frac{\nu}{\kappa},&& \text{and} && \Tilde{\ell}=\frac{\ell}{H} \, ,
\end{align}
where the flux-based Rayleigh number $Ra^{(P)}$ characterizes the strength of the thermal forcing (strength of the imposed radiative heat flux), the Ekman number $E$ characterizes the strength of the global rotation, with $E \ll 1$ for rapid rotation, the Prandtl number $Pr$ characterizes the relative magnitudes of {momentum and thermal diffusivities}, and the dimensionless absorption length $\Tilde{\ell}$ characterizes the spatial structure of the radiative heat source.


We wish to characterize the statistically steady state of the system. Most studies focus primarily on the overall heat transport properties of the system: how large is the overall temperature drop $\Delta T$ that emerges across the fluid layer as a result of the radiative heating? That is, for a given value of the flux-based Rayleigh number $Ra^{(P)}$, one would like to determine the emergent temperature-based Rayleigh number $Ra^{(\Delta T)}$, or, equivalently, the Nusselt number $Nu$, defined as:
\begin{align}
    Ra^{(\Delta T)}=\frac{\alpha g \Delta T H^3}{\nu \kappa}, &&  Nu =\frac{Ra^{(P)}}{Ra^{(\Delta T)}} = \frac{PH}{\rho C \kappa \Delta T} \, .
\end{align}
The goal of the present study is to go beyond the sole quantification of the overall heat transport, with a more in-depth characterization of the temperature and velocity fields: what is the typical flow speed? The typical horizontal scale $\ell^\perp$ of the flow? How do local temperature fluctuations compare to the mean temperature drop? With these questions in mind we introduce the Reynolds number $Re$, the fluctuation-based Rayleigh number $Ra^{(\theta)}$ and the dimensionless horizontal length scale $\ell_*^\perp$, defined as:
\begin{align}
Re = \frac{\sqrt{\left\langle \overline{\mathbf{u}^2} \right\rangle}H}{\nu} \, , \quad Ra^{(\theta)}=\frac{\alpha g  \theta_\mathrm{std}\, H^3}{\nu \kappa} \, , \quad \ell_*^\perp=\frac{\ell^\perp}{H} \, ,
\end{align}
where $\langle\cdot\rangle$ denotes space average, $\overline{\cdot}$ denotes time average {and $\theta_\mathrm{std}=\sqrt{\overline{(\theta-\overline{\theta})^2}}$ denotes the standard deviation of $\theta$ at a given $z$ (see below)}.

\subsection{Laboratory implementation \label{sec:labimp}}

\begin{figure}
    \centering
    \includegraphics[width=8cm]{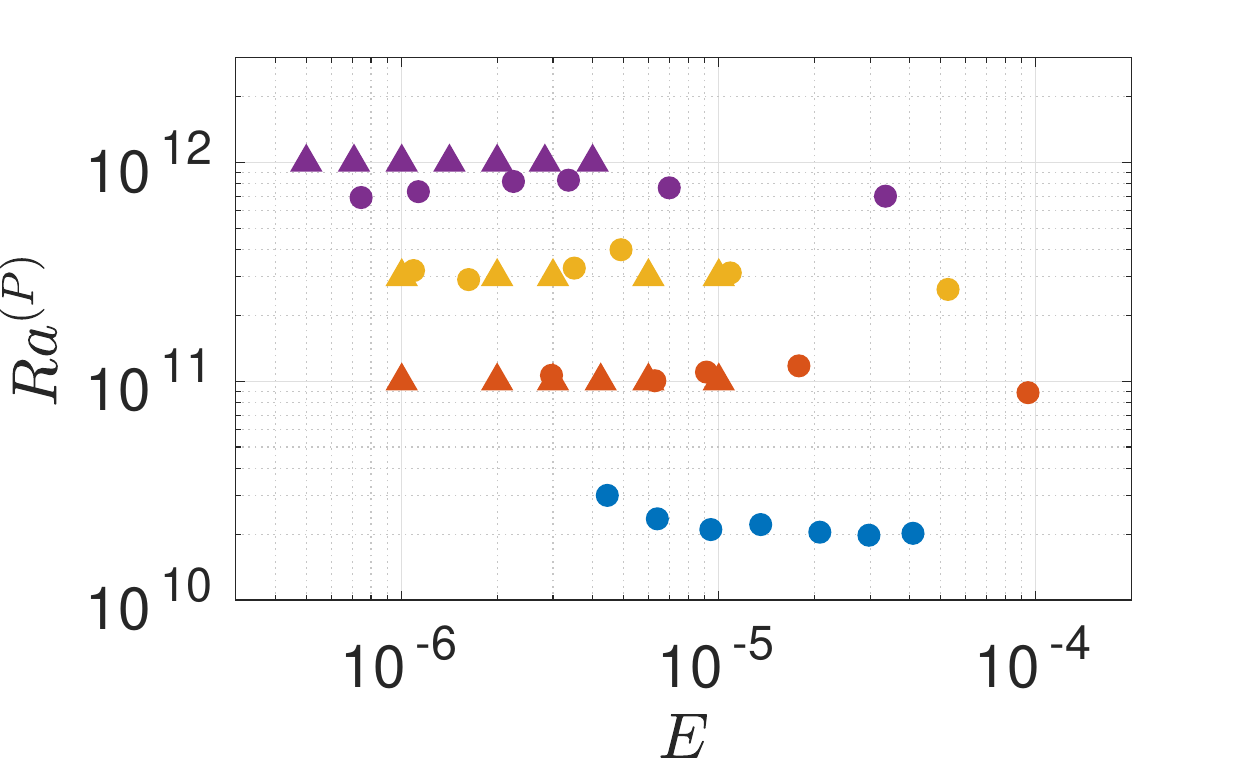}
    \caption{Parameter space spanned by the present dataset. Circles are experimental data ($Pr \approx 7$) while triangles are DNS ($Pr=7$).}
    \label{fig:parameter_space}
\end{figure}

The experimental setup, introduced by~\citet{bouillaut_experimental_2021}, is a rotating version of the radiatively driven convection setup designed by~\citet{lepot_radiative_2018}. A schematic is provided in Figure~\ref{fig:experiment_and_DNS}. A cylindrical tank rotating at a rate $\Omega$ around the vertical axis contains a mixture of water and dye. A powerful spotlight shines at the tank from below. The light passes through a layer of cool water filtering out infrared radiation before reaching the transparent bottom plate of the tank. Absorption of visible light by the dye induces an internal heat source of the form (\ref{eq:defQH}), where the absorption length $\ell$ is directly set by the (uniform) concentration of the dye. The dimensionless absorption length is kept constant in the present study. Specifically, we adopt $\ell=0.048H$ which proves sufficient to bypass the throttling bottom boundary layers and induce diffusivity-free regimes of thermal convection \citep{lepot_radiative_2018,bouillaut_transition_2019,bouillaut_experimental_2021}.

While there is no cooling mechanism in the experimental apparatus, secular heating of the body of fluid amounts to an effective uniform heat sink denoted as $Q_C$ above. Indeed, consider the radiatively heated setup in the absence of cooling. That is, we consider equation (\ref{eq:temperature_eq}) with $Q_C$ set to zero, and we denote the temperature field as $T({\bf x},t)$ instead of $\theta({\bf x},t)$. Space integration over the entire body of fluid, using insulated boundary conditions at all boundaries, yields:
\begin{equation}
    \frac{d \langle T \rangle}{dt}=\frac{P}{\rho CH}\left(1-e^{-H/\ell}\right),
    \label{eq:temperature_drift}
\end{equation}
which indicates that the space-averaged temperature $\langle T \rangle$ increases linearly with time.
This result holds as long as heat losses through the boundaries of the tank are negligible, a valid approximation when $\langle T \rangle$ is within a few degrees of room temperature. 
Now, introduce the variable $\theta({\bf x},t)=T({\bf x},t)-\langle T \rangle(t)$. One can check that the evolution equation for $\theta$ is precisely~(\ref{eq:temperature_eq}), with both the heating term $Q_H$ and the effective cooling term $Q_C$ included. Additionally, the buoyancy force in equation~(\ref{eq:momentum_eq}) can be equivalently cast as $\alpha g T {\bf e}_z$ or as $\alpha g \theta {\bf e}_z$, the difference between the two being absorbed by the pressure gradient. We conclude that equations (\ref{eq:momentum_eq}-\ref{eq:temperature_eq}) indeed model the convective dynamics arising in the laboratory setup.

We leave a free surface at the top of the dyed water, which allows us to easily vary the fluid height $H$ between $10$ and $25$~cm. The second dimensional control parameter is the rotation rate $\Omega$, which we vary between $10$ and $85$~rpm. The corresponding region of the dimensionless parameter space is shown in Figure~\ref{fig:parameter_space}, and we checked in \cite{bouillaut_experimental_2021} that centrifugal effects do not impact the measurements.

We characterize the temperature field using three thermocouples located at height $z=0$, $z=0.25H$ and $z=0.75H$ along the axis of the cylinder. 
An experimental run consists in filling the tank up to a height $H$ with dyed water around $10^o$C, spinning the tank at a given rate $\Omega$ for at least ten minutes to achieve solid-body rotation, and then turning on the spotlight. After some transient, the timeseries of all three probes exhibit a common linear drift at a rate given by the right-hand side of equation~(\ref{eq:temperature_drift}) (examples of timeseries are provided in \cite{bouillaut_experimental_2021}). Following the analysis above, any temperature difference between two probes exhibits a statistically steady signal provided (i) the initial transient phase has decayed and (ii) the fluid temperature is reasonably close to room temperature (typically $\pm 5^o$C). Similarly, temperature fluctuations around the mean drift are statistically steady when conditions (i) and (ii) are met. The experimental data consist of the temperature drop $\Delta T$ measured by \cite{bouillaut_experimental_2021}, which we complement with an estimate of the temperature fluctuations. To wit, we quantify the temperature fluctuations at $z=0.25H$ by subtracting the linear drift from the timeseries before computing the root-mean-square (rms) fluctuations of the resulting statistically steady signal. {This leads to the standard deviation $\theta_\mathrm{std}$ entering the definition of $Ra^{(\theta)}$.}

%

\subsection{Numerical implementation}\label{sec:Num_implementation}

\begin{figure}
    \centering
    \includegraphics[width=14cm]{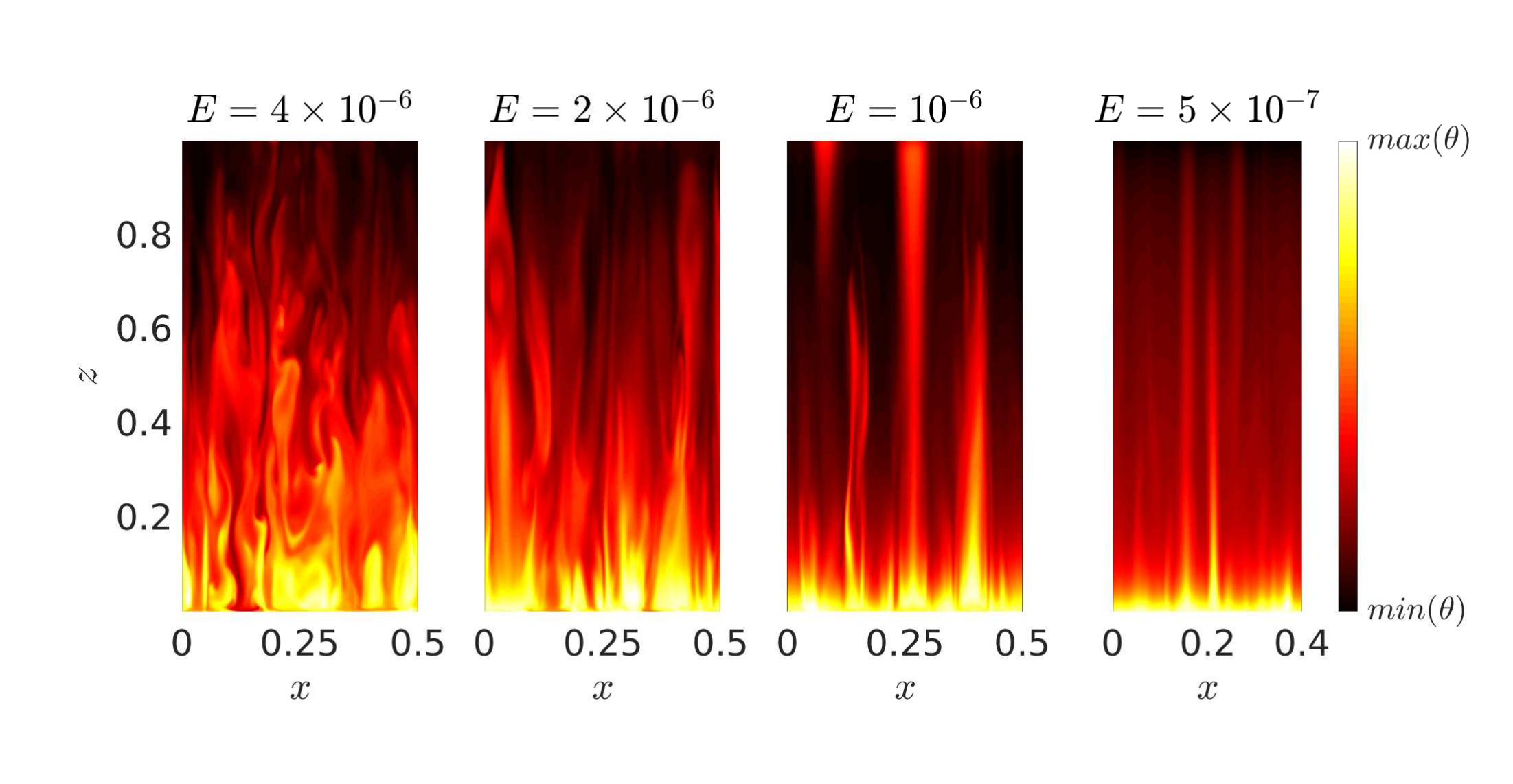}
    \caption{Vertical temperature slices extracted from DNS with $Ra^{(P)}=10^{12}$. {From left to right, the diffusivity-free flux-based Rayleigh number (\ref{eq:defR}) is ${\cal R}=\{1.31\times10^{-6}, 1.63\times10^{-7},2.04\times10^{-8},2.55\times10^{-9}\}$, and the convective Rossby number (\ref{eq:defRo}) is $Ro=\{5.05\times10^{-2}, 2.66\times10^{-2}, 1.77\times10^{-2},1.16\times10^{-2}\} $.} The flow develops thinner columnar structures as the rotation rate increases (left to right).\label{fig:snapshots}}
\end{figure}

To characterize the system beyond the sole quantities accessible in the laboratory experiment, we performed DNS of the governing equations (\ref{eq:momentum_eq}-\ref{eq:temperature_eq}) in a horizontally periodic domain using the pseudo-spectral solver \texttt{Coral}~\citep{miquel_coral_2021}, validated against both analytical results~\citep{miquel_role_2020} and solutions computed with the Dedalus software~\citep{burns_dedalus_2020}. \texttt{Coral} employs a Chebyshev-Fourier-Fourier spatial decomposition and an implicit-explicit time-stepping scheme. In the present study, the upper and lower boundaries are insulated ($\partial_z\theta=0$) and impenetrable ($w=0$). To replicate as closely as possible the experiment, the kinematic boundary conditions are no-slip at the bottom ($u=v=0$) and free-slip at the top ($\partial_z u=\partial_z v = 0$). 

The suite of numerical simulations is focused on the region of parameter space where the experimental data from \cite{bouillaut_experimental_2021} point to the GT scaling regime, see figure~\ref{fig:parameter_space}. The Ekman number varies between $5\times10^{-7}$ and $10^{-4}$, while the flux-based Rayleigh number varies between {$10^{11}$} and $10^{12}$. The Prandtl number is $Pr=7$ for all the DNS. The aspect ratio of the numerical domain is in the range $[0.35,1]$, ensuring that, for each DNS run, at least four wavelengths of the most unstable mode fit along a horizontal direction of the numerical domain. {The typical resolution ranges from $(N_x,N_y,N_z)=(150,150,256)$ for $Ra^{(P)}=10^{11}$ to $(N_x,N_y,N_z)=(300,300,512)$ for $Ra^{(P)}=10^{12}$, where $N_x$ and $N_y$ denote the number of Fourier modes in the horizontal directions, while $N_z$ denotes the number of Chebyschev polynomials in the vertical directions. The non-uniform Chebyschev grid is particularly welcome for the present problem, allowing us to have typically eight grid points inside the Ekman layer after de-aliasing.}

Initial conditions are chosen as either small-amplitude noise, or a solution computed in a previous run. After the initial transient has subsided, we compute the emergent dimensionless parameters of interest by averaging over the statistically steady regime. Because of the periodic boundary conditions, the numerical system is invariant to translations in the horizontal directions. Based on this invariance, when performing a time average $\overline{q}$ of some quantity $q$ extracted from the DNS, we also include a horizontal area average to speed up convergence. Whenever possible, we extract the exact numerical counterpart of the experimental measurements: the temperature difference $\Delta T$ is computed by temporally and horizontally averaging $\theta$ at $z=0$ and $z=0.75H$, before subtracting the two. The fluctuation-based Rayleigh number $Ra^{(\theta)}$ is based on {$\theta_\mathrm{std}$} evaluated at $z=0.25H$. Beyond the experimentally measured quantities, 
DNS gives access to the full statistics of the temperature and velocity fields. We thus also extract the rms velocity $\la \overline{{\bf u}^2} \ra^{1/2}$ over the entire fluid domain to compute the emergent Reynolds number $Re$. Finally, we extract a characteristic horizontal length scale $\ell^\perp$ of the flow as: 
\begin{equation}
    \ell^\perp=\left( \frac{\left\langle \overline{\psi^2} \right\rangle}{\left\langle \overline{ \mathbf{u}^2} \right\rangle} \right)^{1/2} \, , \label{eq:defellperp}
\end{equation}
where $\psi(x,y,z,t)$ denotes the so-called toroidal streamfunction, defined as minus the inverse horizontal Laplacian of the vertical vorticity {(for a purely horizontal flow, this definition leads to the standard streamfunction of the 2D flow in a given constant-$z$ plane, see e.g. in \cite{julien2007reduced} for details on the toroidal/poloidal decomposition).}

In figure~\ref{fig:snapshots}, we show vertical slices of the temperature field for fixed $Ra^{(P)}=10^{12}$ and decreasing values of the Ekman number (increasing rotation rate). One observes thinner structures with strong vertical coherence as the rotation rate increases. {As compared with standard rotating Rayleigh-Bénard convection, the top-down asymmetry of the present system is visible in the temperature snapshots. This asymmetry arises predominantly as a result of the asymmetry of the distribution of heat sources and sinks, and to a lesser extent as a result of the asymmetry between the (free-slip) top and (no-slip) bottom boundary conditions. One observes stronger vertical temperature variations in the vicinity of the heating region, with a quieter region in the upper half of the domain. The temperature fluctuations measured by the probe located at $z=0.75H$ are thus weaker than the fluctuations measured at $z=0.25H$, although we checked that they display the same scaling behavior with the control parameters of the system. More generally, and anticipating the results in the next sections, the good agreement between the scaling-laws measured in the present setup and in the idealized top-down-symmetric setup of \cite{barker_theory_2014} indicates that the top-down asymmetry does not impact the  scaling behavior of the various quantities of interest in the GT regime.}

\section{The geostrophic turbulence regime: theoretical background \label{sec:theory_scalings}} 

\subsection{Asymptotically fast rotation, $E\ll 1$}

As discussed at the outset, deriving scaling predictions for the various emergent quantities of interest proves more challenging for rotating convection than for non-rotating convection, because of the additional dimensionless parameter $E$ in the former case. Progress can be made in the rapidly rotating regime $E \ll 1$, as initially proposed by \cite{stevenson_turbulent_1979} and put on firm analytical footing by \cite{julien_new_1998} through an asymptotic expansion of the equations in powers of $E^{1/3}$ \citep{aurnou_connections_2020}. As for any asymptotic expansion, a challenging part of the analysis consists in inferring the correct scaling of the various quantities of interest with the small parameter $E^{1/3}$. We propose here a recipe to determine these scalings based on the following observation: the reduced set of equations for rapid rotation must hold near the threshold of instability. That is, the sought scalings can be inferred from the structure of the most unstable eigenmode arising near the threshold for instability. This approach provides a shortcut to deriving the predictions of the geostrophic turbulence scaling regime. The first step consists in writing the scaling in $E$ of the various quantities characterizing the most unstable eigenmode computed through linear stability analysis (see \cite{chandrasekhar_hydrodynamic_1961} for rotating Rayleigh-B\'enard convection and \cite{bouillaut_radiative_2022} for the rotating radiatively driven setup). In the rapidly rotating regime $E \ll 1$, the marginal state is characterized by the following asymptotic relations:
\begin{subequations}
\label{eq:linrels}
\begin{eqnarray}
 \begin{array}{r}
      \text{Threshold temperature-based }\\ 
      \text{Rayleigh number:} 
 \end{array} 
 & Ra^{(\Delta T)}_c \sim E^{-4/3}\, , & \label{eq:linrel1} \\[0.2cm]
 {\text{Growthrate (or angular frequency):}} & {\sigma \sim \frac{\kappa}{H^2} E^{-2/3}}\, , \label{eq:growthrate} & \\[0.2cm]
 \text{Horizontal lengthscale:} & \ell^{\perp} \sim H E^{1/3}\, , & \\[0.2cm]
 \text{Vertical lengthscale:} &  \ell_z \sim H\, , & \\[0.2cm]
 \begin{array}{r}
      \text{Ratio of temperature to velocity}\\
      \text{fluctuations:}
 \end{array}
 & {\theta_\mathrm{std}} \sim \frac{\Delta T H}{\kappa} \la \overline{w^2} \ra^{1/2} E^{2/3}\, , &  \qquad \\[0.2cm]
 \begin{array}{r}
      \text{Relation between the various velocity}\\ 
      \text{components:} 
 \end{array} 
 & \la \overline{u^2} \ra \sim \la \overline{v^2} \ra \sim \la \overline{w^2} \ra \, . & \label{eq:linrel5}
\end{eqnarray}
\end{subequations}
Based on the asymptotic behavior of the temperature-based Rayleigh number, one introduces the reduced Rayleigh number $\widetilde{Ra}=Ra^{(\Delta T)} E^{4/3}$. We focus on the rapidly rotating near-threshold regime corresponding to the distinguished limit $E \to 0$ with fixed $\widetilde{Ra}={\cal O}(E^0)$. {Equation (\ref{eq:growthrate}) above indicates the behavior of the growthrate of the most unstable mode in this distinguished limit, or simply the behavior of the frequency of oscillation at threshold when convection arises through a Hopf bifurcation~\citep{chandrasekhar_hydrodynamic_1961}. More generally, in the distinguished limit of interest here, all the $\sim$ symbols in equations (\ref{eq:linrel1}-\ref{eq:linrel5}) can be replaced by an equals sign, at the expense of multiplying the right-hand side by a generic function ${\cal F}(\widetilde{Ra},Pr)$ (in the following, the symbol ${\cal F}$ denotes a generic functional dependence that {\it a priori} differs between successive equations). Indeed, at the instability threshold, $\widetilde{Ra}$ is constant and one must recover the scalings (\ref{eq:linrel1}-\ref{eq:linrel5}) in Ekman number.}

{The saturation level of the various fields is determined based on the dominant nonlinearity of the equations~\citep{stevenson_turbulent_1979}. Here the nonlinearities  are all of advective type, entering the expression of the total derivative as:}
\begin{equation}
\partial_t + (\mathbf{u}\cdot \bm{\nabla}) = \partial_t + \mathbf{u}_\perp \cdot \bm{\nabla}_\perp + w \partial_z \, .
\end{equation}
{In the linear regime of the instability (exponential growth), the term $\partial_t$ is of the order of the growth rate $\sigma$, while the advective nonlinearities are negligible. Saturation arises when these advective nonlinearities become comparable to $\sigma$. One easily checks using (\ref{eq:linrel1}-\ref{eq:linrel5}) that vertical advection is negligible as compared to horizontal advection:}
\begin{equation}
w \partial_z \sim \frac{w}{H} \sim \frac{u_\perp}{H} \sim \frac{\ell^\perp}{H} \times \frac{u_\perp}{\ell^\perp} \sim E^{1/3} \, \mathbf{u}_\perp \cdot \bm{\nabla}_\perp \ll \mathbf{u}_\perp \cdot \bm{\nabla}_\perp \, .
\end{equation}
{Saturation thus arises when $\mathbf{u}_\perp \cdot \bm{\nabla}_\perp \sim \la \overline{{\bf u}^2} \ra^{1/2}/\ell^\perp$ becomes comparable to $\sigma$. Substitution of the scalings (\ref{eq:linrel1}-\ref{eq:linrel5}) yields the scaling for the velocity at saturation:}
\begin{equation}
\la \overline{{\bf u}^2} \ra^{1/2} = \frac{\kappa}{H} E^{-1/3} {\cal F}(\widetilde{Ra},Pr) \, , \label{eq:scalingu}
\end{equation}
{which we recast as a scaling for the Reynolds number:}
\begin{equation}
\frac{\la \overline{w^2} \ra^{1/2} H}{\nu} \sim \frac{\la \overline{{\bf u}^2} \ra^{1/2} H}{\nu}  =  E^{-1/3} {\cal F}(\widetilde{Ra},Pr)\, . \label{eq:salingtempRe}
\end{equation}

%

With the heat flux $P/(\rho C)$ scaling as the convective flux $\la \overline{w \theta} \ra \sim \la \overline{w^2} \ra^{1/2} \theta_\mathrm{std}$, one can deduce the scaling in Ekman number of all the quantities of interest listed above at finite distance from threshold based on the combination of {(\ref{eq:scalingu})} with (\ref{eq:linrels}):
\begin{subequations}
\label{sys:3.4}
\begin{align}
Nu & =  {\cal F}(\widetilde{Ra},Pr) \, , \\
\frac{\ell^\perp}{H} & =  E^{1/3} {\cal F}(\widetilde{Ra},Pr)\, ,  \label{tempellperp}\\
\frac{\ell_z}{H} & =  {\cal F}(\widetilde{Ra},Pr)\, , \\
\frac{\theta_\mathrm{std}}{\Delta T} & =  E^{1/3} {\cal F}(\widetilde{Ra},Pr) \, , \label{tempthrms} 
\end{align}
\end{subequations}
where, again, the symbol ${\cal F}$ denotes a generic function that differs between successive equations. The equations above provide the dominant scalings in Ekman number behind the derivation of the reduced model by \cite{julien_new_1998}.
\subsection{Diffusivity-free regime}
In practical terms, the distinguished limit considered above allows one to consider increasingly large values of $\widetilde{Ra}$ as the Ekman number $E$ decreases. 
For extremely low values of $E$, one can even consider values of $\widetilde{Ra}$ that are much greater than one. The leap of faith is then to assume that there is a regime of low-enough Ekman number for {(\ref{eq:salingtempRe}-\ref{sys:3.4})} to hold, but far enough from threshold  ($\widetilde{Ra} \gg 1$) to reach a diffusivity-free scaling regime (numerical integration of the associated reduced model does provide evidence for this, see e.g. \cite{julien_heat_2012}). The diffusivity-free scaling argument then amounts to demanding that the functions ${\cal F}$ in {(\ref{eq:salingtempRe})} through (\ref{tempthrms}) be such that $\kappa$ and $\nu$ can be crossed out from both sides of the equations, which finally leads to the GT scaling predictions:
\begin{eqnarray}
Nu \sim \frac{\widetilde{Ra}^{3/2}}{Pr^{1/2}}& \sim & \left(Ra^{(\Delta T)}\right)^{3/2} E^2 Pr^{-1/2}\, , \\
\frac{\ell^\perp}{H} \sim E^{1/3} \sqrt{\frac{\widetilde{Ra}}{Pr}}  & \sim & E \sqrt{\frac{Ra^{(\Delta T)}}{Pr}}\, , \label{sec1:pred2}\\
\frac{\ell_z}{H} & \sim & 1\, , \\
\frac{\la {\bf u}^2 \ra^{1/2} H}{\nu} \sim E^{-1/3} \frac{\widetilde{Ra}}{Pr}& \sim & \frac{Ra^{(\Delta T)} \, E}{Pr}\, , \\
\frac{\theta_\mathrm{std}}{\Delta T} \sim E^{1/3} \sqrt{\frac{\widetilde{Ra}}{Pr}} & \sim & E \sqrt{\frac{Ra^{(\Delta T)}}{Pr}} \, . \label{sec1:pred5}
\end{eqnarray}
To highlight the diffusivity-free form of these predictions, we combine the emergent parameters of interest with $E$ and $Pr$ to form dimensionless combinations that do not involve the molecular diffusivities $\kappa$ and $\nu$. For radiatively driven convection, the central control parameter is then the diffusivity-free flux-based Rayleigh number ${\cal R}$, defined as:
\begin{eqnarray}
{\cal R}= \frac{Ra^{(P)} E^3}{Pr^2} \, . \label{eq:defR}
\end{eqnarray}
Denoting with a star the diffusivity-free form of the emergent dimensionless parameters, we introduce:
\begin{eqnarray}
Nu_*  =  \frac{Nu E}{Pr} \, , \quad Ra^{(\theta)}_*=\frac{Ra^{(\theta)} E^2}{Pr} \, , \quad Re_*=E\, Re \, ,  \quad \ell_*^\perp=\frac{\ell^\perp}{H} \, .
\end{eqnarray}
Such diffusivity-free dimensionless numbers allow us to recast the GT scaling predictions in a particularly compact form reported in Table~\ref{tab:predictions}. Namely, each diffusivity-free emergent parameter evolves as some power-law in ${\cal R}$ with a specific exponent.

\begin{table}
  \begin{center}
\def~{\hphantom{0}}
  \begin{tabular}{lllc}
      \multicolumn{1}{c}{Dimensional quantity}  &  \multicolumn{1}{c}{Dimensionless parameter} & \multicolumn{1}{c}{Diffusivity-free form}  & \multicolumn{1}{c}{GT prediction} \\[3pt]
      \hline
       $\Delta T$ 									& $Nu=\frac{PH}{\rho C \kappa \Delta T}$  				& $Nu_*=EPr^{-1}Nu$  & $Nu_*\sim \mathcal{R}^\frac{3}{5}$\\
       $\theta_\mathrm{std}$ 				& $Ra^{(\theta)}=\frac{\alpha g  \theta_\mathrm{std} H^3}{\kappa \nu}$ 		& $Ra^{(\theta)}_*=E^2Pr^{-1}Ra^{(\theta)}$  & $Ra^{(\theta)}_*\sim \mathcal{R}^\frac{3}{5}$\\
       ${\la\overline{{\bf u}^2}\ra}^{1/2}$ 				& $Re = \frac{{\la \overline{\mathbf{u}^2} \ra}^{1/2} H}{\nu}$ 				& $Re_*=E\, Re$ &  $Re_*\sim \mathcal{R}^\frac{2}{5}$\\
       $\ell^\perp$  								& $\frac{\ell^\perp}{H}$ 		& $\ell_*^\perp = \frac{\ell^\perp}{H}$ &  $\ell^\perp_*\sim \mathcal{R}^\frac{1}{5}$\\
  \end{tabular}
  \caption{Scaling predictions for the various emergent quantities of interest in the geostrophic turbulence (GT) scaling regime, expressed in terms of the diffusivity-free flux-based Rayleigh number ${\cal R}= {Ra^{(P)} E^3}/{Pr^2}$.   \label{tab:predictions}}
  \end{center}
\end{table}

\section{Experimental and numerical assessment of the scaling predictions}

\begin{figure}
    \centering
    \includegraphics[width=9cm]{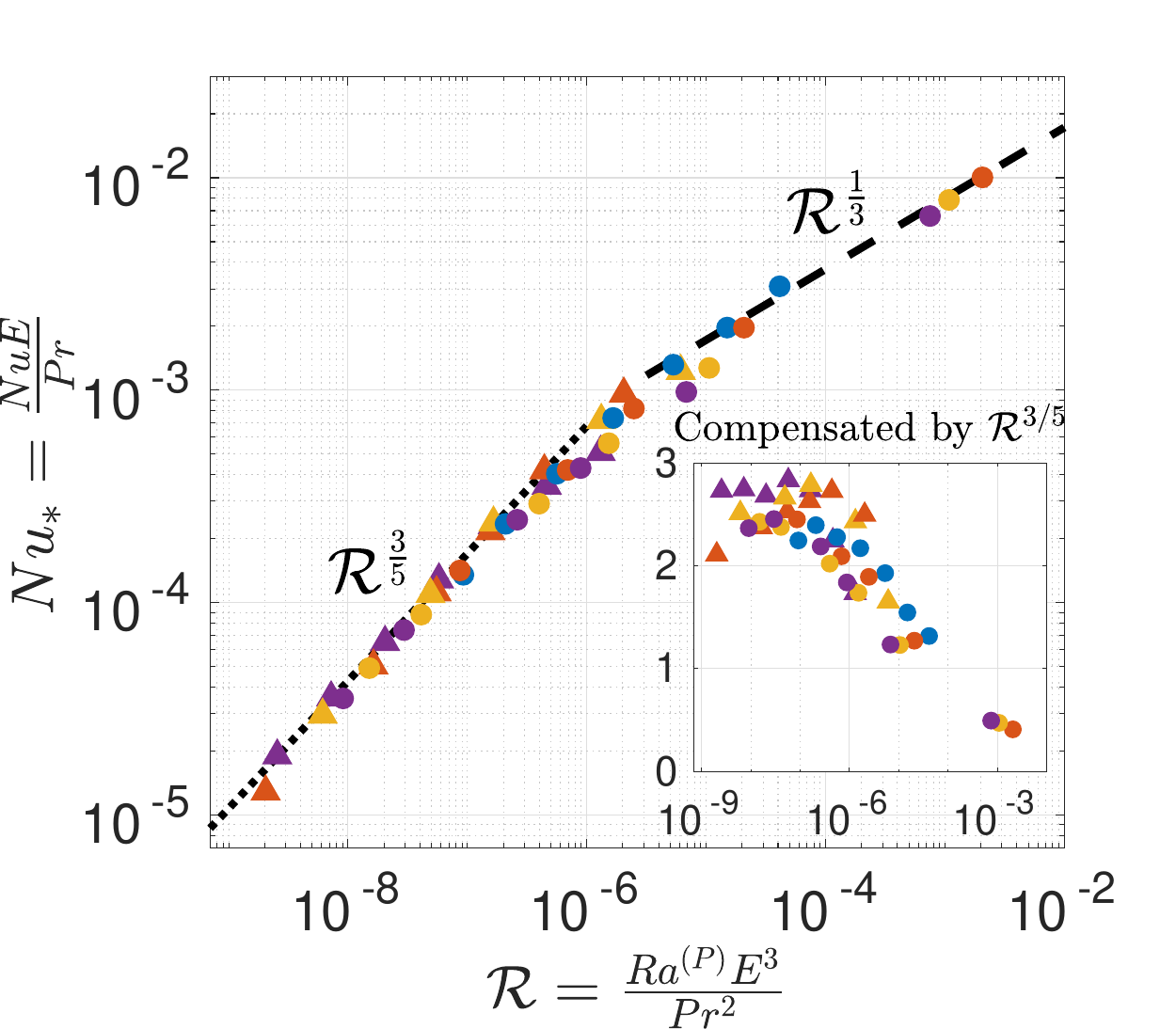}
    \caption{Diffusivity-free Nusselt number $Nu_*$ as a function of the diffusivity-free flux-based Rayleigh number ${\cal R}$. Same symbols as in figure \ref{fig:parameter_space}. The dotted line indicates the GT scaling exponent $3/5$. The dashed line is the prediction of the non-rotating `ultimate' regime, characterized by an exponent of $1/3$ in this representation. The inset shows $Nu_*/{\cal R}^{3/5}$  versus ${\cal R}$.}
    \label{fig:Nusselt}
\end{figure}

\subsection{Heat transport}

As discussed at the outset, a central question in turbulent convection is the scaling relation between the heat flux and the overall temperature drop. The GT scaling prediction for the heat transport has been previously validated in \cite{bouillaut_experimental_2021} by plotting the experimental data for $Nu_*$ as a function of ${\cal R}$. We reproduce this plot in Figure~\ref{fig:Nusselt} for completeness, to which we add the numerical data. We confirm the excellent agreement between the experimental data and the GT scaling prediction $Nu_* \sim {\cal R}^{3/5}$ for ${\cal R} \leq 3\times 10^{-7}$. The numerical data are also in excellent agreement with the prediction. The compensated plot further highlights this agreement and indicates that the prefactors of the scaling-laws inferred from the experimental and numerical data are fully compatible, the numerical prefactor being greater by approximately $15$\%.

\subsection{Temperature fluctuations}

\begin{figure}
    \centering
    \includegraphics[width=9cm]{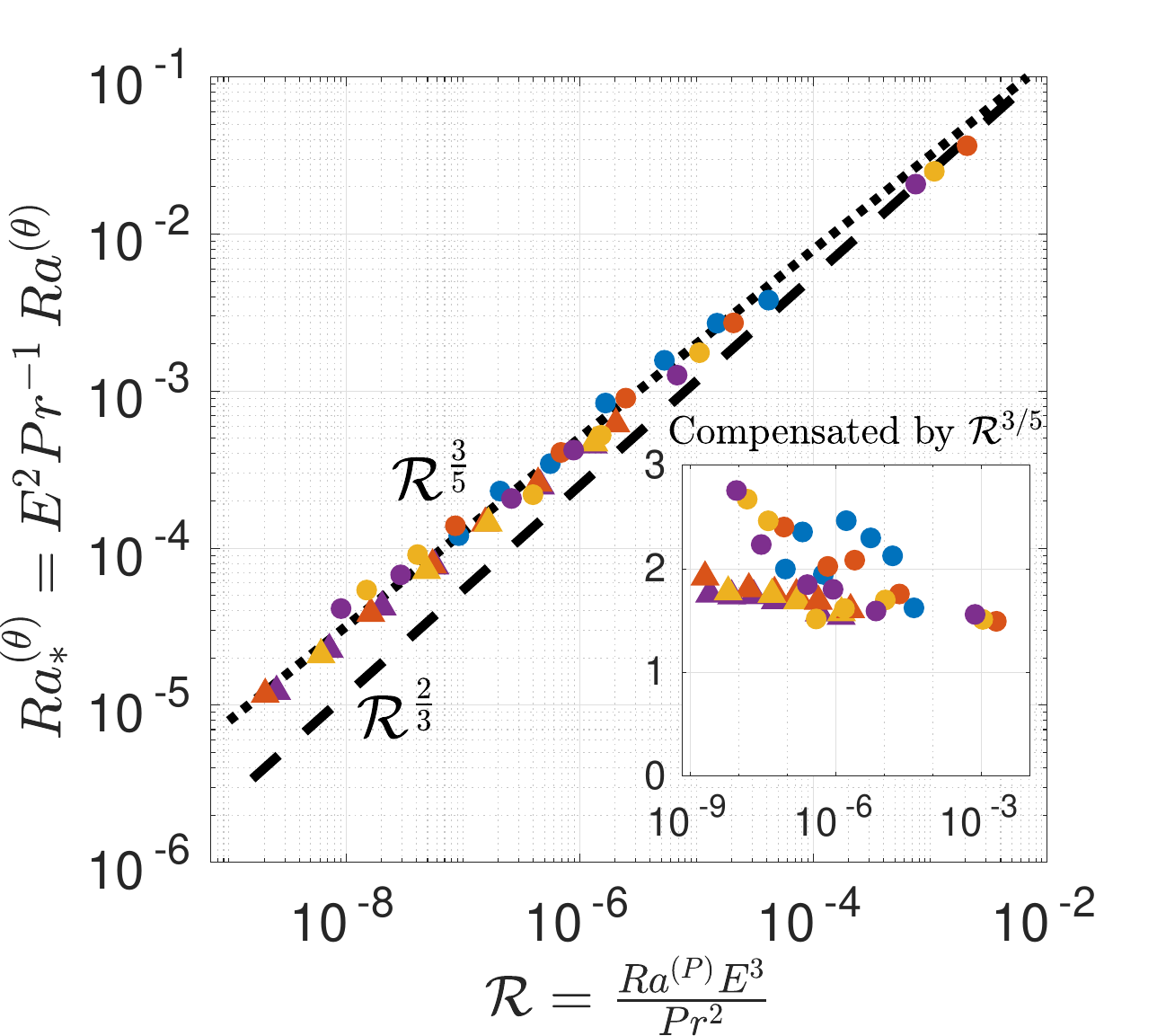}
    \caption{Diffusivity-free fluctuation-based Rayleigh number $Ra^{(\theta)}_*$ as a function of ${\cal R}$. Same symbols as in Figure~\ref{fig:parameter_space}. The dotted line indicates the GT scaling prediction {${\cal R}^{3/5}$, while the dashed line indicates the scaling prediction ${\cal R}^{2/3}$ of the non-rotating `ultimate' scaling regime}. Inset: $Ra^{(\theta)}_*/{\cal R}^{3/5}$ versus ${\cal R}$.}
    \label{fig:Ra_theta}
\end{figure}

We now turn to the temperature fluctuations, which we also measure both experimentally and numerically. To assess the validity of the GT scaling prediction, in Figure~\ref{fig:Ra_theta}, we plot the diffusivity-free fluctuation-based Rayleigh number $Ra^{(\theta)}_*$ as a function of ${\cal R}$. Again, this representation leads to a very good collapse of the data, indicating diffusivity-free (or `ultimate') dynamics. There is arguably more scatter in the experimental data than in the numerical data, because the latter benefit from an area horizontal average together with the time average when computing the rms temperature fluctuations (see section \ref{sec:Num_implementation}). Still, the theoretical prediction ${\cal R}^{3/5}$ is validated over four decades in ${\cal R}$, with a prefactor that is greater by approximately 20\% for the experimental data than for the numerical data. Somewhat surprisingly, the range of validity of the GT prediction seems to extend beyond ${\cal R} = 3\times 10^{-7}$, that is, the prediction for $Ra^{(\theta)}_*$ is validated over a more extended range of ${\cal R}$ than the prediction for $Nu_*$.


\subsection{Convective flow speed}

\begin{figure}
    \centering
    \includegraphics[width=9cm]{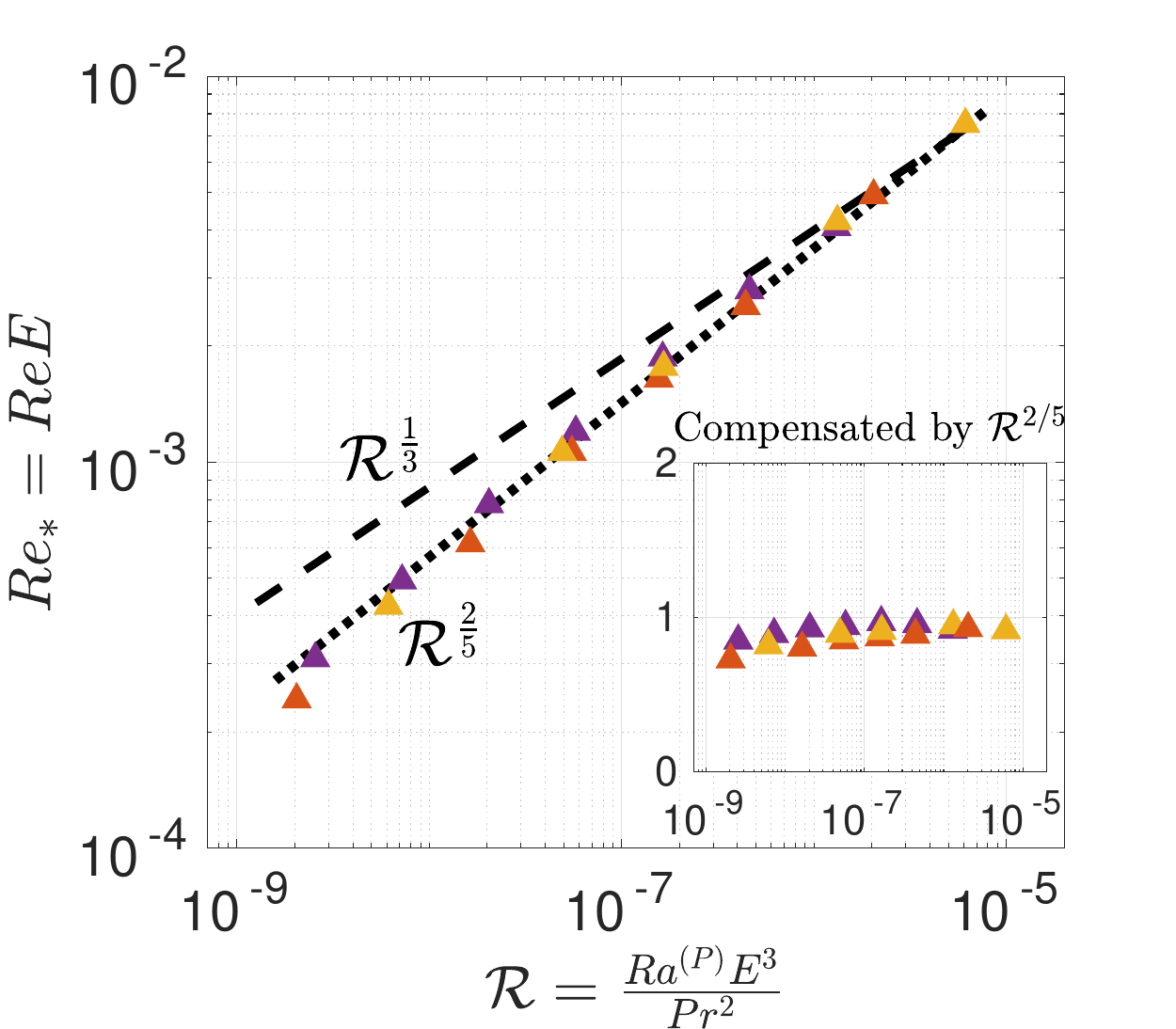}
    \caption{Diffusivity-free Reynolds number $Re_*$ extracted from the DNS as a function of ${\cal R}$. Same symbols as in Figure~\ref{fig:parameter_space}. The dotted line indicates the GT scaling prediction ${\cal R}^{2/5}$, while the dashed line indicates the scaling prediction ${\cal R}^{1/3}$ of the non-rotating `ultimate' scaling regime. Inset: $Re_*/{\cal R}^{2/5}$ versus ${\cal R}$.    \label{fig:Reynolds}}
\end{figure}

As compared with the laboratory experiment, DNS allows us to readily characterize the velocity field. We thus turn to the rms velocity in the fluid domain inferred from the numerical simulations. In Figure~\ref{fig:Reynolds}, we plot the diffusivity-free Reynolds number $Re_*$ as a function of ${\cal R}$. Once again, we obtain an excellent collapse of the numerical data onto a single master curve, which indicates diffusivity-free dynamics. The master curve is in excellent agreement with the GT prediction ${\cal R}^{2/5}$, as shown by the eye-guide in the main figure and by the compensated plot in the inset. Also shown in the figure is the prediction ${\cal R}^{1/3}$ associated with the non-rotating ultimate regime of thermal convection \citep{bouillaut_velocity-informed_2022}, which is incompatible with the present rapidly rotating data points. We also note that the GT scaling prediction for $Re_*$ appears to be valid over a more extended range of ${\cal R}$ than the prediction for $Nu_*$. The reason may be that the exponents $2/5$ and $1/3$ of the rotating and non-rotating scaling predictions are rather close, possibly inducing a very smooth and extended crossover region between the two power-law behaviors.



\subsection{Horizontal lengthscale}

\begin{figure}
    \centering
    \includegraphics[width=9cm]{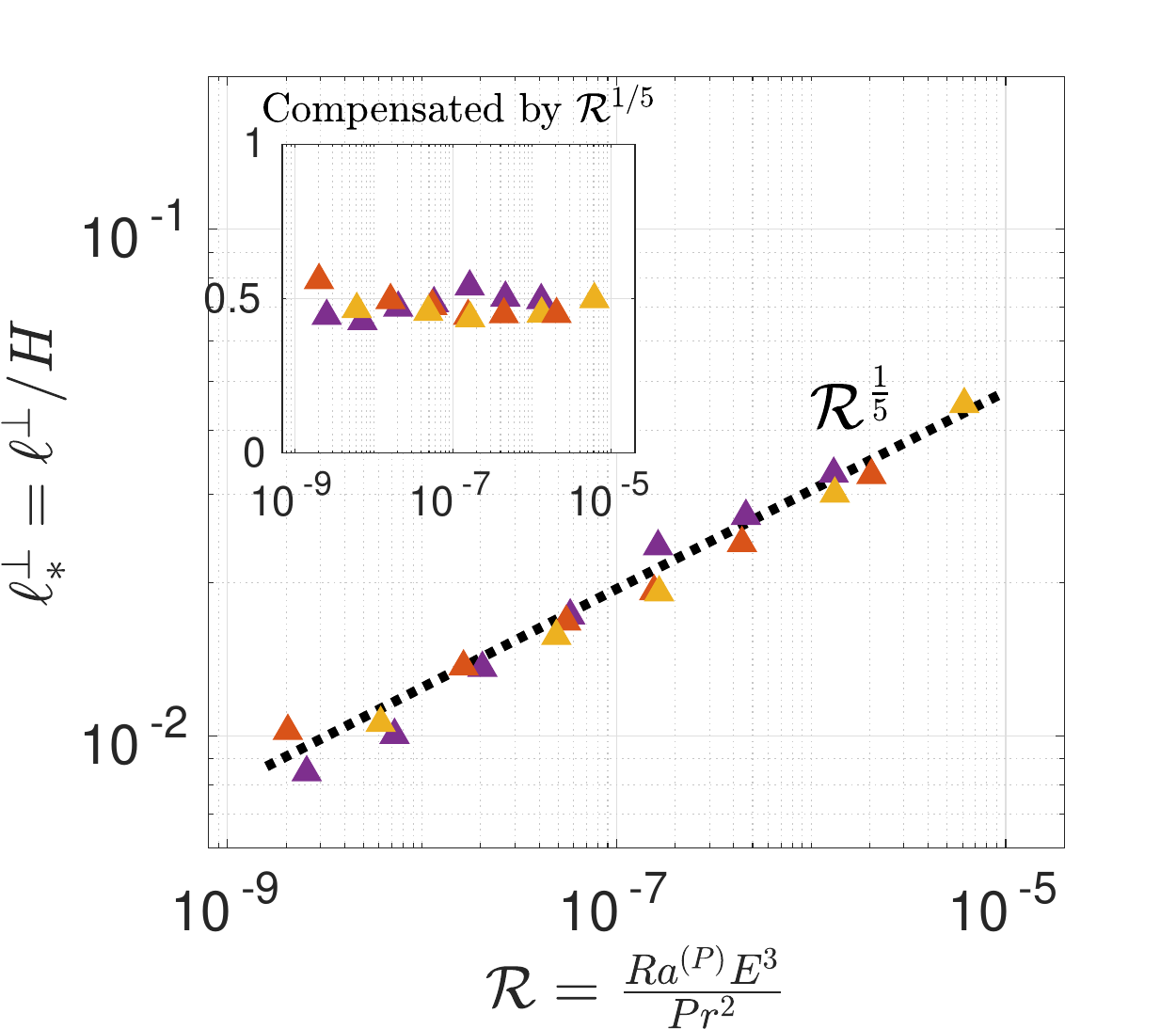}
    \caption{Dimensionless horizontal scale $\ell_*^\perp$ as a function of ${\cal R}$. The dotted line indicates the GT scaling prediction. Inset: $\ell_*^\perp/{\cal R}^{1/5}$ versus ${\cal R}$.}
    \label{fig:ell_perp}
\end{figure}

To further characterize the convective flow we now turn to its characteristic horizontal lengthscale, extracted from the DNS based on the definition~(\ref{eq:defellperp}). In Figure~\ref{fig:ell_perp}, we plot $\ell_*^\perp$ as a function of ${\cal R}$ for the numerical dataset. Once again, this representation leads to a good collapse of the data, although the scatter appears greater than for the previous quantities of interest. This is likely a consequence of the shallower GT scaling prediction ${\cal R}^{1/5}$ for this quantity of interest. The eyeguide in the main figure indicates good agreement with this prediction, which is further confirmed by the compensated plot in the inset. 

{Comments are in order regarding the definition (\ref{eq:defellperp}) of the horizontal lengthscale of the flow. Indeed, previous studies have reported different scaling behaviors using different proxies for the horizontal scale of the flow, only a subset of which agree with the GT prediction~\citep{de2023tidal,oliver2023small}. One reason is that many of these proxies directly involve the dissipative scales and thus the viscosity. For instance, one should refrain from using the Taylor scale or the Kolmogorov scale of the flow when investigating GT scaling, as these scales are (partly) controlled by the viscous scale. Similarly, one should refrain from using the rms vorticity of the flow to diagnose the horizontal flow scale, as the rms vorticity is typically controlled by the small viscous scales. In other words, one expects to observe the zeroth law of turbulence (and therefore the GT predictions) for large-scale quantities, such as the horizontal integral scale of the flow defined in equation (\ref{eq:defellperp}). As discussed in the following section, the no-slip bottom boundary condition of the present setup prevents the emergence of domain-scale vortices, such that $\ell^\perp$ in (\ref{eq:defellperp}) is always smaller than the horizontal extent of the domain. This seems to be a simplification as compared with studies employing stress-free boundary conditions.}

\section{{Discussion}}

Based on a hybrid experimental-numerical study, we have validated the scaling predictions of the GT regime for rapidly rotating radiatively driven convection. {The scaling predictions for the heat transport and temperature fluctuations are validated using both experimental and numerical data. Following such cross-validation, the scaling predictions for the flow speed and horizontal scale are validated through further processing of the numerical data.}

The GT scaling predictions for the temperature fluctuations, the flow speed and the horizontal scale of the flow are observed over many decades in diffusivity-free flux-based Rayleigh number ${\cal R}$, and they seem to arise even before the Nusselt number displays GT scaling. 
The same observation seems to hold for the rotating Rayleigh-B\'enard system, where numerical studies indicate that the observation of the GT prediction for the heat transport requires more extreme parameter values than the GT predictions for the flow speed and flow structures \citep{julien_heat_2012,song2024scaling}. At the experimental level, \cite{vogt_oscillatory_2021} report velocity measurements that are compatible with the GT scaling prediction, even though the GT scaling regime for the heat transport remains elusive in rotating Rayleigh-B\'enard experiments. Similarly, \cite{guervilly_turbulent_2019} report diffusivity-free behavior for the Reynolds number and the azimuthal scale of the flow in a numerical study  of rotating convection in a spherical shell, while the Nusselt number has still to reach the GT regime. These observations confirm that GT scaling for the Nusselt number arises further into the asymptotic regime of control parameters than GT scaling for the flow scale and flow speed. {Although the diffusivity-free flux-based Rayleigh number ${\cal R}$ is a natural control parameter of our experiment, regime transitions in the Rayleigh-Bénard setup are more often characterized in terms of the convective Rossby number:}
\begin{equation}
Ro = \sqrt{\frac{\alpha g \Delta T}{4 H \Omega^2}} = \sqrt{\frac{\cal R}{Nu_*}} \, , \label{eq:defRo}
\end{equation}
{the rapidly rotating regime corresponding typically to $Ro \lesssim 0.1$. We provide the values of $Ro$ for each panel of figure~\ref{fig:snapshots} in the figure caption to illustrate the transition to rapidly rotating convection. The GT asymptote in figure~\ref{fig:Nusselt} corresponds to $Nu_*\simeq 2.8 \, {\cal R}^{3/5}$, which after substitution in (\ref{eq:defRo}) yields $Ro=0.6 \, {\cal R}^{1/5}$. As discussed above, clear GT scaling for the Nusselt number is observed for ${\cal R} \lesssim 3\times 10^{-7}$. The corresponding transition convective Rossby number is $Ro=0.6 \, (3\times 10^{-7})^{1/5} = 0.03$, within the expected range.}

{As mentioned in the introduction, the idealized numerical study that is closest to the present setup is that of~\cite{barker_theory_2014}. The authors employ a combination of local internal heating and cooling localized near the boundaries, together with an imposed background temperature gradient that is relaxed to a target value, with the goal of ensuring no average flux through the fixed-temperature top and bottom boundaries (the background temperature gradient is fixed in some simulations, while it obeys a relaxation equation when fluctuations are too strong). With this approach the authors successfully validate the scaling predictions of \cite{stevenson_turbulent_1979}, provided the heat transport is characterized using the temperature gradient at mid-depth and the Reynolds number is based on the vertical velocity component. Our setup can be seen as an experimental realization of such internally heated and cooled convection, further confirming the numerical results of \cite{barker_theory_2014} while avoiding the subtleties of their setup and diagnostics. For instance, the boundaries of the laboratory experiment are thermally insulated to a good approximation, without the need for the adjusted background temperature gradient of \cite{barker_theory_2014}. Similarly, our DNS employs Chebyshev expansions to readily account for such insulated boundaries (contrasting with the Fourier decomposition in \cite{barker_theory_2014}). Additionally, we follow the standard practice of defining the Nusselt number based on a temperature drop, showing consistent behavior with the temperature gradient considered by \cite{barker_theory_2014}.
Finally, we believe that the solid boundary at the bottom of the experimental tank -- and associated no-slip boundary condition in the DNS -- also contributes to `simplifying' the setup, in the sense that it prevents (or delays) any accumulation of horizontal kinetic energy into large-scale vortices~\citep{favier2014inverse,guervilly2014large,stellmach_approaching_2014,kunnen_transition_2016,favier2019subcritical,aguirre2020competition,maffei2021inverse}.  (Indeed, running a few numerical test cases with a free-slip bottom boundary condition did lead to large-scale-vortex formation sufficiently far from threshold, while large-scale vortices were not observed for the realistic no-slip bottom boundary condition. When they develop, these large scale vortices modify the scaling for the horizontal scale and the horizontal speed of the flow.) Overall, our study shows excellent agreement with the pioneering study of \cite{barker_theory_2014} while offering a `realistic' setup, in the sense that it can be realized in the laboratory.}

To conclude, radiatively driven convection offers a robust way of achieving the diffusivity-free (or ultimate) regimes of thermal convection in the laboratory, with and without global rotation. Additional physical ingredients could be included in the experimental and numerical setups, with the goal of investigating the diffusivity-free regimes of a broader range of natural convective flows. Magnetic field crucially influences convection in planetary and stellar interiors and could be the natural next step.

\backsection[Acknowledgements]{This work was granted access to the HPC resources of TGCC under the allocation A0122A10803, A0142A10803 and A0142A12489 made by GENCI.}

\backsection[Funding]{This research is supported by the European Research Council under grant agreement FLAVE 757239.}

\backsection[Declaration of interests] {The authors report no conflict of interest.}


\backsection[Author ORCIDs]{\\
G. Hadjerci, https://orcid.org/0009-0008-4707-5217;\\
B. Miquel, https://orcid.org/0000-0001-6283-0382;\\ 
B. Gallet, https://orcid.org/0000-0002-4366-3889.}



\end{document}